\documentclass[prb,twocolumn,longbibliography,
 aps,floatfix,nofootinbib]{revtex4-2}

\usepackage{graphicx}
\usepackage{dcolumn}
\usepackage{bm}
\usepackage{amssymb}
\usepackage{amsmath}
\usepackage{float}
\usepackage[colorlinks=true, citecolor=blue, urlcolor=blue]{hyperref}

\usepackage{epstopdf}
\AppendGraphicsExtensions{.tif}

\newcommand{\be}{\begin{equation}}
\newcommand{\ee}{\end{equation}}
\newcommand{\tjn}{T_\textrm{JN}}
\newcommand{\gth}{G_\textrm{th}}
\newcommand{\kh}{\hat{\kappa}}
\newcommand{\sh}{\hat{\sigma}}
\newcommand{\grad}{\vec{\nabla}}
\newcommand{\rr}{\vec{r}}
\newcommand{\qq}{\vec{q}}

\begin{document}

\preprint{APS/123-QED}

\title{Relation between Johnson noise and heating power in a two-terminal conductor}

\author{Calvin Pozderac and Brian Skinner\\
\it{Department of Physics, Ohio State University, Columbus, Ohio 43210, USA}}

\date{\today}

\begin{abstract}

We consider the Johnson noise of a two-dimensional, two-terminal electrical conductor for which the electron system obeys the Wiedemann-Franz law. We derive two simple and generic relations between the Johnson noise temperature and the heat flux into the electron system. First, we consider the case where the electron system is heated by Joule heating from a DC current, and we show that there is a universal proportionality coefficient between the Joule power and the increase in Johnson noise temperature. Second, we consider the case where heat flows into the sample from an external source, and we derive a simple relation between the Johnson noise temperature and the heat flux across the boundary of the sample.

\end{abstract}

\maketitle

\section{Introduction}
\label{sec:introduction}

Electrical current flowing through a resistor at finite temperature exhibits random temporal fluctuations known as Johnson-Nyquist noise \cite{johnson_thermal_1928, nyquist_thermal_1928}. One can think that this noise arises because thermal fluctuations in the distribution of electron velocities act like random current sources within the sample. The net effect of these random current sources is to produce a current noise $\Delta I$ whose mean-square value is proportional to the electron temperature $T_e$ as
\be 
\langle(\Delta I)^2 \rangle  = \frac{4 k_\textrm{B} T_e}{R} \Delta f .
\label{eq:dI2}
\ee 
Here, $k_\textrm{B}$ is the Boltzmann constant, $R$ is the sample resistance, and $\Delta f$ is the noise bandwidth. For a given measurement of $\langle(\Delta I)^2 \rangle$, Eq.~(\ref{eq:dI2}) defines the ``Johnson noise temperature''.\footnote[2]{Equation (\ref{eq:dI2}) can be viewed as a result of the fluctuation-dissipation theorem at zero frequency and the Wiener-Khinchin theorem. It is only valid for samples with purely resistive impedence and at frequencies $\lesssim k_B T/h$, where $h$ is Planck's constant \cite{nyquist_thermal_1928}.}

Electron noise thermometry is a powerful experimental technique which exploits this relation between current noise and electron temperature in order to produce an accurate measurement of the electron temperature \cite{fong_ultrasensitive_2012, crossno_development_2015, mckitterick_ultrasensitive_2015, qu_johnson_2019, talanov_high-bandwidth_2020}. This technique has proven especially fruitful as a method for making ultra-sensitive bolometers using graphene electrons \cite{karasik_prospective_2014, efetov_fast_2018, miao_graphene-based_2018, liu_towards_2018, miao_demonstration_2021}, or for more fundamental studies of the electron thermal conductivity and heat capacity \cite{fong_measurement_2013, yigen_electronic_2013, yigen_wiedemannfranz_2014, crossno_observation_2016, talanov_high-bandwidth_2020, waissman_measurement_2021}.
A very recent work \cite{waissman_measurement_2021} has developed a nonlocal thermometry technique, in which heat flows across a ``bridge'' material of interest and into an electron system (monolayer graphene). The authors (including one of us) showed that, by measuring the increase in the Johnson noise in the graphene, one can infer the heat flow across the bridge, and thus measure its thermal conductance. 

\begin{figure}[tb!]
\begin{center}
\includegraphics[width=0.95\columnwidth]{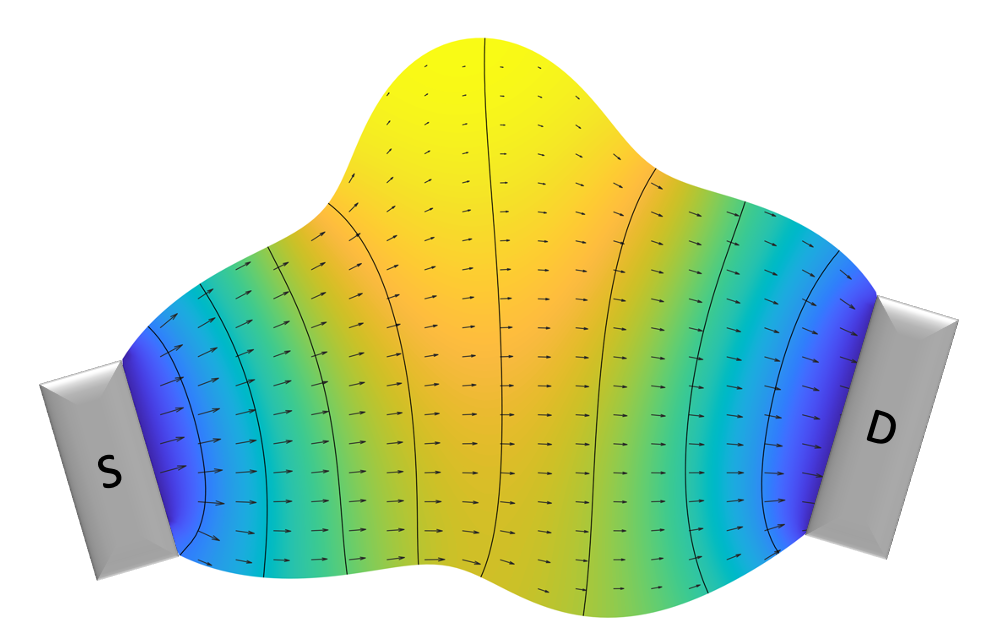}
\end{center}
\caption{A schematic illustration of a two-terminal conductor with Johnson noise measured between source (S) and drain (D) contacts. Contour lines indicate contours of the characteristic potential $\phi$, and arrows indicate $-\grad \phi$. The color corresponds to a hypothetical distribution of electron temperature, produced either by Joule heating or by injection of heat along the lateral edges, with colder temperatures near the contacts.}
\label{fig:flow}
\end{figure}

Importantly, however, in all of these experimental contexts the electron temperature is generally nonuniform across the sample. For example, the metal contacts tend to act as good heat sinks, keeping the edges of the sample at the lower bath temperature while Joule heating or other extraneous heat sources cause the electron temperature to increase toward the interior of the sample (as depicted in Fig.~\ref{fig:flow}).

Making use of electron noise thermometry therefore requires one to understand the relationship between the Johnson noise, as measured between a given pair of contacts, and the distribution of electron temperature across the sample. Previous theoretical work has examined this relationship in some generality, including the case where there are many electrodes with different boundary conditions for the electrical current \cite{sukhorukov_noise_1999, song_shockley-ramo_2014,Kadomtsev_1957,Kogan_1969,Blanter_2000,Heikkila_2013}. The purpose of the present paper is to focus on the two-terminal setup, and to derive a set of general relationships between the measured Johnson noise temperature and the heating power that may be readily used in a variety of experiments.

Our primary motivation is the experimental setup of Ref.~\cite{waissman_measurement_2021}, in which Johnson noise measurements serve as a tool for measuring thermal conductance. Previous experiments have examined the situation where the sample is heated via resistive Joule heating, allowing one to infer the thermal conductance of the electron system by observing the corresponding increase in Johnson noise temperature  \cite{fong_ultrasensitive_2012, crossno_development_2015, qu_johnson_2019, talanov_high-bandwidth_2020, fong_measurement_2013, yigen_electronic_2013, yigen_wiedemannfranz_2014, crossno_observation_2016}. Below we examine this situation in detail, and show that there is a simple and universal relation between the Joule power and the increase in Johnson noise temperature. More interestingly, we also consider the setup pioneered by Ref.~\cite{waissman_measurement_2021}, in which heat flows into the electron system via a ``bridge'' material, whose properties may be unknown. The relation that we derive between the heating power injected from the bridge and the increase in Johnson noise temperature of the electron system enables one to infer the thermal conductance of the bridge. In this way the technique provides a powerful new way of studying a variety of emergent excitations within the bridge material \cite{waissman_measurement_2021}.

Throughout this paper we consider situations where electrical current flows between a single pair of source and drain contacts. We focus on the case where the electrical and thermal conductivity tensors, $\sh$ and $\kh$, respectively, obey the Wiedemann-Franz law,
\be 
\kh = L_0 T_0 \sh,
\label{eq:WF}
\ee 
where $L_0 = (\pi^2/3)(k_\textrm{B}/e)^2$ is the Wiedemann-Franz ratio, with $-e$ the electron charge, and $T_0$ is the base electron temperature.
We further restrict ourselves to the case where the electron temperature $T_e$ deviates only slightly from $T_0$, and to the ``hot-electron'' limit, in which electron-electron collisions are strong enough to locally equilibrate electrons among themselves, while electron-phonon collisions and other sources of inelastic scattering are negligible. The validity of neglecting the contributions from phonons, particularly in graphene, has been discussed at length in previous work \cite{Efetov_e_ph,Kubakaddi_e_ph,Bistritzer_e_ph,Tse_e_ph,Viljas_e_ph,Song_e_ph}.
Within this set of assumptions, we derive two simple relationships between the Johnson noise temperature and the heating power applied to the electron system, which we summarize here before deriving below. 

First, we consider the case where all heating to the electron system is provided by a DC voltage $V$ that is applied between the source and drain contacts (as in Refs.~\onlinecite{crossno_development_2015, fong_ultrasensitive_2012, fong_measurement_2013, yigen_wiedemannfranz_2014}). In this situation we show that 
\be 
\tjn = \frac{P R}{12 L_0 T_0},
\label{eq:TJNintro}
\ee
where $R$ is the two-terminal resistance and $P = V^2/R$ is the Joule power dissipated in the system (and thus $\tjn$ is independent of $R$). Here and below, $\tjn$ denotes the increase in Johnson noise temperature from the base temperature $T_0$.
Equation (\ref{eq:TJNintro}) has been derived previously for the special case of a rectangular sample with spatially uniform current (e.g., in Refs.~\onlinecite{fong_measurement_2013, yigen_electronic_2013}). But here we show that it holds generically for any two-terminal setup, regardless of geometry or of the structure and spatial variation of the conductivity tensor, so long as the local electric and thermal currents are described by Ohm's law and the Wiedemann-Franz law.
(As we discuss in more detail below, Eq.~(\ref{eq:TJNintro}) can also be seen as a limiting case of a more general expression derived in Ref.~\cite{sukhorukov_noise_1999}.)

Second, we consider the case where there is no Joule heating, but heat is injected into the system through its boundaries. In this case, the value of $\tjn$ depends only on a boundary integral of the incoming heat current density multiplied by a characteristic potential that vanishes at either contact [see Eq.~(\ref{eq:tjnedgefinal}) below].
In the special case of interest in Ref.~\onlinecite{waissman_measurement_2021}, where $\kh$ is diagonal and the heat power $Q$ is injected along a line of bilateral symmetry in the system,
\be 
\tjn = \frac{Q R}{8 L_0 T_0}.
\label{eq:tjnsymmetric}
\ee

\section{Relation between $\tjn$ and electron temperature}
\label{setup}

We consider a two-terminal setup, in which the current is measured between source and drain electrodes connected to a conducting sample with arbitrary shape and arbitrary conductivity tensor $\sh$ (which need not be diagonal). Within the interior of the sample, the current density $\vec{j}$ is related to the electric potential $\Phi$ by Ohm's law,
\be 
\vec{j} = - \sh \vec{\nabla} \Phi
\ee 
(i.e., we assume linear, Ohmic response of the current everywhere.)
The continuity equation for the current within the sample (the Laplace equation) is $\grad \cdot \vec{j} = 0$, or
\be 
\vec{\nabla} \cdot (\sh \vec{\nabla} \Phi) = 0.
\label{eq:Laplace}
\ee 
The boundaries of the sample that do not coincide with either of the two contacts have a no-current condition,
\be 
(\sh \grad \Phi) \cdot \hat{n} = 0,
\label{eq:phiBC}
\ee
where $\hat{n}$ is an (outward-facing) unit normal vector.

The local heat current $\qq(\rr)$ at position $\rr$ is related to the electron temperature $T(\rr)$ by the heat equation
\be 
\qq = - \kh \grad T.
\ee 
For convenience, we define the local electron temperature $T(\rr) = T_e(\rr) - T_0$ relative to the base temperature $T_0$, and we consider the limit where $T \ll T_0$, so that Eq.~(\ref{eq:WF}) remains valid.
In regions where the electron system is being heated, the local power density $p(\rr)$ injected into the electron system satisfies
\be 
p = - \grad \cdot (\kh \grad T)
\label{eq:heat}
\ee 
(the heat equation). Throughout this paper we assume that all heat current is carried by electrons (i.e., we neglect the phonon contribution to $\kh$ and we ignore electron-phonon scattering). We also assume that the contacts act as good heat sinks that are held at the base temperature, so that both contacts constitute $T = 0$ boundary conditions.

The Johnson noise is related to the local temperature $T(\rr)$ via a characteristic potential $\phi(\rr)$, which relates the intensity of a local current source to the magnitude of current collected between the source and drain electrodes. Specifically \cite{sukhorukov_noise_1999, song_shockley-ramo_2014},
\be
    \tjn = \frac{\int d^2 r T(\rr) \grad \phi(\rr) \cdot [\sh \grad \phi(\rr)]} { \int d^2 r \grad \phi(\rr) \cdot [\sh \grad \phi(\rr)] }.
\label{eq:TJNdef}
\ee 
In other words, the quantity $\grad \phi(\rr) \cdot [\sh \grad \phi(\rr)]$ acts as a weighting function for the local electron temperature.

The characteristic potential $\phi(\rr)$ satisfies the same Laplace equation and lateral boundary conditions as the true potential $\Phi(\rr)$ [Eqs.~(\ref{eq:Laplace}) and (\ref{eq:phiBC})]. We emphasize, however, that the characteristic potential is independent of any actual voltage applied between the source and drain, and $\phi(\rr)$ is well-defined even in situations where no voltage is applied and $\Phi(\rr) \equiv 0$ everywhere. Without loss of generality, one can choose the normalization of $\phi$ such that $\phi = 1$ at the source contact and $\phi = 0$ at the drain. We will show below that $\tjn$ is independent of the choice of labels for the two contacts, i.e.\ that $\tjn$ is unchanged by the operation $\phi \rightarrow 1 - \phi$. 

With this choice of normalization for $\phi$, the denominator of Eq.~(\ref{eq:TJNdef}) represents the Joule power dissipated when a unit voltage is applied between the two contacts, which is equal to the inverse of the two-terminal resistance $R$:
\be 
\int d^2 r \grad \phi(\rr) \cdot [\sh \grad \phi(\rr)] = \frac{1}{R}.
\label{eq:R}
\ee 
Hence, we can rewrite Eq.~(\ref{eq:TJNdef}) as
\be 
    \tjn = R \int d^2 r T(\rr) \grad \phi(\rr) \cdot [\sh \grad \phi(\rr)].
    \label{eq:TJNR}
\ee

Equation (\ref{eq:TJNR}) is true generically, regardless of how the temperature distribution $T(\rr)$ is established.

\section{Heating by DC current}
\label{sec:Joule}

In this section we consider the case where the electron system is heated via Joule heating by a DC voltage source (``self-heating''), and there is no heat flow across the lateral boundaries of the sample. The total Joule power absorbed by the electron system is equal to $P = V^2/R$, where $V$ is the voltage between the two contacts. The Joule heating produces a temperature distribution that peaks in the interior of the sample, as the electrons conduct the dissipated Joule heat to the contacts (as depicted in Fig.~\ref{fig:flow}).

If we define the electric potential such that $\Phi = V$ at the source and $\Phi = 0$ at the drain, then the true electric potential $\Phi(\rr)$ and the characteristic potential $\phi(\rr)$ are directly proportional to each other, $\Phi(\rr) = V \phi(\rr)$.  The local Joule power density $p_\textrm{J} = \vec{j} \cdot \vec{E}$, where $\vec{E} = - \grad \Phi$ is the electric field, or 
\be 
p_\textrm{J} = V^2 (\sh \grad \phi) \cdot (\grad \phi).
\label{eq:pJ}
\ee 
(Here and below we suppress the argument $\rr$ of $\phi$.)
Equating the Joule power $p_\textrm{J}$ to the right-hand side of the heat equation [Eq.~(\ref{eq:heat})], and using the Wiedemann-Franz law, gives
\be 
    V^2 ( \sh \grad \phi) \cdot (\grad \phi) = 
    -L_0 T_0 \grad \cdot (\sh \grad T).
    \label{eq:intermediate1}
\ee 

From this equation we can derive a relation between the temperature distribution $T(\rr)$ and the characteristic potential $\phi(\rr)$. First, notice that ${(\sh \grad \phi) \cdot (\grad \phi) = \grad \cdot (\phi \sh \grad \phi) - \phi \grad \cdot (\sh \grad \phi)}$ (the chain rule). The second term in this expression is zero by the Laplace equation [Eq.~(\ref{eq:Laplace})], so that Eq.~(\ref{eq:intermediate1}) can be rewritten as
\be 
 - \frac{V^2}{L_0 T_0} \grad \cdot (\phi \sh \grad \phi) = 
    \grad \cdot (\sh \grad T).
    \label{eq:intermediate2}
\ee 
This equation and the relevant boundary conditions (${T = 0}$ at both contacts) is satisfied uniquely by the temperature distribution
\be 
T(\rr) = \frac{V^2}{2 L_0 T_0} \phi(\rr) \left( 1 - \phi(\rr) \right).
\label{eq:Tphi}
\ee 
Notice, as mentioned above, that the temperature distribution is invariant under the relabelling of the two contacts, $\phi \rightarrow 1 - \phi$. Notice also that the heat current across the lateral boundaries ${-(\kh \grad T) \cdot \hat{n} \propto (1 - 2 \phi)(\sh \grad \phi)\cdot \hat{n} = 0}$, so that the appropriate boundary conditions are satisfied.

We can now manipulate Eq.~(\ref{eq:TJNR}) for the Johnson noise temperature in order to understand its relation with the Joule power $P$. On the one hand, a direct substitution of Eq.~(\ref{eq:Tphi}) for $T(\rr)$ into the definition of $\tjn$ [Eq.~(\ref{eq:TJNR})] gives
\be 
\tjn = \frac{V^2 R}{2 L_0 T_0} \int d^2 r \phi ( 1 - \phi) \grad \phi \cdot (\sh \grad \phi).
\label{eq:tjnphi1}
\ee 

On the other hand, one can arrive at an equivalent expression for $\tjn$ by equating the Joule power $p_\textrm{J}$ with the divergence of the heat current [Eq.~(\ref{eq:heat})], and substituting the resulting expression for $\grad \phi \cdot (\sh \grad \phi)$ into the definition of $\tjn$ [Eq.~(\ref{eq:TJNR})]. This process gives
\be 
\tjn = - \frac{R}{V^2} \int d^2 r T(\rr) \grad \cdot (\kh \grad T).
\ee
Integrating this expression by parts (using Green's theorem), and making use of the fact that either $T = 0$ or $(\kh \grad T) \cdot \hat{n} = 0$ along all the boundaries of the sample, we find
\be 
\tjn = \frac{R}{V^2} \int d^2 r (\grad T) \cdot (\kh \grad T).
\ee
Plugging in the expression for $T$ in terms of $\phi$ [Eq.~(\ref{eq:Tphi})] and using the Wiedemann-Franz law gives
\begin{eqnarray} 
\tjn & = & \frac{V^2 R}{4 L_0 T_0} \int d^2 r (1 - 2 \phi)^2 (\grad \phi) \cdot (\sh \grad \phi) \nonumber \\
 & = & \frac{V^2 R}{4 L_0 T_0} \times \left[ -4 \int d^2 r \phi (1 - \phi) (\grad \phi) \cdot (\sh \grad \phi) \nonumber \right. \\
 & & + \left. \int d^2 r (\grad \phi) \cdot (\sh \grad \phi)  \right] \nonumber \\
 & = & -2 \tjn + \frac{V^2}{4 L_0 T_0}. 
\end{eqnarray}
The last line of this sequence follows from the expression for $\tjn$ in Eq.~(\ref{eq:tjnphi1}) and from the expression for the two-terminal resistance in Eq.~(\ref{eq:R}). Thus we arrive at the final, simple and generic result
\be 
\label{TJN1}
\tjn = \frac{V^2}{12 L_0 T_0},
\ee 
which is equivalent to Eq.~(\ref{eq:TJNintro}) announced in the Introduction. We note that Eq.~(\ref{TJN1}) can be obtained by taking the weak heating limit of a more general expression presented in Eq.~(5.5) of Ref.~\cite{sukhorukov_noise_1999}. For convenience, we briefly re-derive that more general expression in Appendix~\ref{Appendix}. It is also interesting to note that the Johnson noise temperature has a simple relation with the maximum temperature in the sample, $T_\text{max}$. This relation can be seen by maximizing Eq.~(\ref{eq:Tphi}) with respect to $\phi$, which gives $\tjn = 2T_\text{max}/3$ \cite{Naidyuk_T_max}.

If one defines the thermal conductance $\gth$ such that $P = \gth \tjn$, then
\be 
\gth = 12 L_0 T_0/R.
\label{eq:gth}
\ee 
This expression for $\gth$ has been used in a number of experimental works (e.g., Refs.~\onlinecite{fong_measurement_2013, yigen_electronic_2013, yigen_wiedemannfranz_2014, crossno_development_2015, crossno_observation_2016, talanov_high-bandwidth_2020}), and was derived for the special case of a rectangular sample with uniform electric current. But here we have shown that it is completely generic, independent of both the geometry of the sample and the form of the conductivity tensor. Equation (\ref{eq:gth}) may break down only if the Wiedemann-Franz law is violated or if Ohm's law ceases to hold (due, for example, to the formation of quantum Hall edges states, or to a mean free path that is longer than some geometric dimension of the sample).

\section{Heating by injected heat current}
\label{sec:injected}

In this section we consider the case where there is no significant Joule heating in the sample, and instead the electron system is heated by an injection of heat current along the lateral boundaries of the sample (as in the experimental setup of Ref.~\onlinecite{waissman_measurement_2021}).

In this situation without Joule heating, the heat current is conserved throughout the interior of the sample, ${\grad \cdot (\kh \grad T) = 0}$, while the heat current flowing across the lateral boundaries $-(\kh \grad T) \cdot \hat{n}$ need not be zero. The characteristic potential, however, still satisfies $(\sh \grad \phi)\cdot \hat{n} = 0$ along the lateral boundaries, since it describes the continuity of electrical current.

From the definition of $\tjn$ in Eq.~(\ref{eq:TJNR}) and the Wiedemann-Franz law,
\be 
\tjn = \frac{R}{L_0 T_0} \int d^2 r T(\rr) \grad \phi(\rr) \cdot [\kh \grad \phi(\rr)].
\ee 
We can integrate this expression by parts using two applications of Green's first identity, together with the condition that either $(\sh \grad \phi) \cdot \hat{n} = 0$ or $T = 0$ along the boundaries of the sample. This procedure gives
\be 
\tjn = \frac{R}{L_0 T_0} \int d^2r (1 - \phi) (\grad T) \cdot (\kh \grad \phi).
\ee 
We can now use the vector identity ${\vec{X} \cdot (\hat{M}\vec{Y}) = (\hat{M}^T\vec{X}) \cdot \vec{Y}}$ to arrive at
\be 
\tjn = \frac{R}{L_0 T_0} \int d^2r (1 - \phi) (\grad \phi) \cdot (\kh^T \grad T).
\ee
Using one more integration by parts, together with the conditions $\phi ( 1 - \phi) = 0$ at the terminals and ${\grad \cdot (\kh^T \grad T) = \grad \cdot (\kh \grad T) = 0}$ in the interior, we arrive at
\be 
\tjn = \frac{R}{2 L_0 T_0} \int_C ds \, \phi (1 - \phi) (-\kh^T \grad T) \cdot (-\hat{n}).
\label{eq:tjnedgefinal}
\ee 
Here the notation $\int_C ds$ denotes a contour integral around the boundaries of the sample. Notice, as above, that the expression for $\tjn$ depends on $\phi$ only through the combination $\phi(1-\phi)$, so that it is independent of the choice of labels for the two contacts.

In the case where $\kh$ is symmetric, so that $\kh = \kh^T$, Eq.~(\ref{eq:tjnedgefinal}) can be interpreted simply as
\be 
\tjn = \frac{R}{2 L_0 T_0} \int_C ds \, \phi (1 - \phi) \qq \cdot (-\hat{n}),
\ee 
where the term $\qq \cdot (-\hat{n})$ denotes the heat current density that is injected into the system along the sample boundary. So, for example, when heat is injected close to one of the contacts, where $\phi ( 1 - \phi)$ is small, the corresponding increase $\tjn$ in the Johnson noise temperature is small. One can think that $\tjn$ is small in this case because the injected heat is absorbed immediately by the nearest contact without providing much heating of the electron system. The largest value of $\tjn$ for a given heat flux is realized when the heat is injected at a point along a line of bilateral symmetry in the system, such that $\phi = 1 - \phi = 1/2$. In this special case (relevant for the experiments of Ref.~\onlinecite{waissman_measurement_2021}),
\be 
\tjn = \frac{Q R}{8 L_0 T_0},
\ee 
where $Q$ is the injected heat power.









\section{Summary and Conclusion}

In this paper we have derived the relationship between the Johnson noise temperature and the heating power for two generic situations that are relevant for Johnson noise thermometry. In the case of ``self-heating'' setups, where heating to the electron system is provided by a DC current, our primary result is Eq.~(\ref{eq:TJNintro}) [or, equivalently, Eq.~(\ref{eq:gth})]. This result has been used in a number of experiments, as derived for a rectangular sample with diagonal conductivity tensor and spatially uniform current \cite{fong_measurement_2013, yigen_electronic_2013, yigen_wiedemannfranz_2014, crossno_development_2015, crossno_observation_2016, talanov_high-bandwidth_2020}. But it is not generally appreciated that the result holds generically for any geometry and even in the presence of a magnetic field or other source of Hall conductivity. The breakdown of Eq.~(\ref{eq:TJNintro}) implies either a breakdown of Ohm's law (as may arise, for example, from the formation of quantum Hall edge states), or the breakdown of the Wiedemann-Franz law (as may arise from electron-phonon coupling). 

We have also examined the situation where the electron system is heated by an injection of heat current along the boundary, as in the nonlocal thermometry setup of Ref.\ \onlinecite{waissman_measurement_2021}. The most generic result for $\tjn$ in this setup is Eq.\ (\ref{eq:tjnedgefinal}), which has a simple interpretation in terms of the injected heat current density when the thermal conductivity tensor is diagonal. In the case where heat is injected at a point along a line of bilateral symmetry, $\tjn$ adopts the simple form of Eq.~(\ref{eq:tjnsymmetric}). This limiting case result is given a simplified derivation in the Supplementary Information of Ref.~\onlinecite{waissman_measurement_2021}.

The approach we have presented can in principle be generalized to the case of more than two contacts, although we have not attempted to do so. In this case there is a separate characteristic potential $\phi_{nm}$ for each pairs of contacts $n$, $m$ \cite{sukhorukov_noise_1999}. One should also be careful about the boundary conditions associated with other contacts (i.e., whether they are grounded or floating), which has an effect on the characteristic potential. We will note, however, that the two-contact description is appropriate for describing the Johnson noise measured between any pair of contacts that are relatively well separated from all others. Indeed, if the electrical conductance between contacts $n$ and $m$ is much larger than the conductance between either $n$ or $m$ and any other contact, then the Johnson noise between $n$ and $m$ can be well-approximated by the two-contact description used here.

\acknowledgments

We thank Jonah Waissmann and Philip Kim for many helpful discussions and for a related collaboration.

\def\bibsection{\section*{\refname}} 

\bibliography{TJN.bib}

\clearpage

\appendix
\section{\label{Appendix}General $\tjn$ Expression from Joule Heating}

In Sec.~\ref{sec:Joule} we consider Johnson noise resulting from Joule heating by a DC voltage source in the limit of small deviations of the electron temperature $T$ from the base electron temperature $T_0$. In fact, this calculation can be carried out more generally for any temperature deviation (so long as the assumptions of Ohm's law, the Wiedemann-Franz law, and the hot electron regime remain justified). Such a calculation has been performed in Ref.~\cite{sukhorukov_noise_1999}. Here, for convenience, we present a brief re-derivation of this more general result.

Starting with Eq.~(\ref{eq:intermediate1}), but keeping the Wiedemann-Franz law as $\kh =L_0 (T_0 + T)\sh$, we have:
\begin{equation}
\label{laplace}
    V^2(\sh\grad\phi)\cdot(\grad\phi) = -L_0 \grad\cdot((T_0+T)\sh\grad T).
\end{equation}
Through the use of the continuity equation, this expression is written as
\begin{equation}
    V^2\grad\cdot(\sh \grad (\phi (1- \phi))) = L_0 \grad\cdot(\sh\grad ((T_0+T)^2)).
\end{equation}

From this expression, one can make the ansatz that $V^2\grad (\phi(1-\phi)) = L_0\grad((T_0+T)^2)$. By making use of the boundary conditions [$\phi(1-\phi)=T=0$ at the contacts], we arrive at
\begin{equation}
   T(\phi) = \sqrt{T_0^2 + \frac{V^2}{L_0}\phi(1-\phi)}-T_0.
\end{equation}

Plugging this expression back into Eq.~(\ref{laplace}) validates our ansatz. Further, if we take the limit of small $V$, we reproduce Eq.~(\ref{eq:Tphi}). 

In order to calculate the Johnson temperature in the more general case, we first derive the following relationship:
\begin{equation}
\label{identity}
    \int d^2r T(\phi)(\grad\phi)\cdot(\sh\grad\phi) = \frac{1}{R} \int_0^1 T(\phi)d\phi.
\end{equation}

We know $f'(\phi)(\grad\phi)\cdot(\sh\grad\phi) = (\grad f(\phi))\cdot(\sh\grad\phi) = \grad\cdot(f(\phi)\sh\grad\phi)$. Therefore, this equality along with an application of Green's theorem yields
\begin{equation}
\label{inter}
    \int d^2r f'(\phi)(\grad\phi)\cdot(\sh\grad\phi) = \int ds f(\phi)\sh\grad\phi\cdot \hat{n}.
\end{equation}

Since there is no electric current flowing across the lateral boundaries, $\sh\grad\phi\cdot \hat{n}$ is only non-zero at the contacts. Further, $1/R=\int d^2r (\grad \phi)\cdot(\sh\grad\phi)= \int ds_{(\phi=1)} \sh\grad\phi\cdot \hat{n}$. Since the total current out of the drain is equal to the current into the source,  $\int ds_{(\phi=0)}\sh\grad\phi\cdot \hat{n} = -1/R$. Combining these expressions with Eq.~(\ref{inter}) results in
\begin{align}
\begin{split}
    \int d^2r f'(\phi)(\grad\phi)\cdot(\sh\grad\phi) &=\frac{1}{R}(f(1)-f(0))\\
    &=\frac{1}{R}\int_0^1 f'(\phi) d\phi.
\end{split}
\end{align}

By letting $f'(\phi) = T(\phi)$ we arrive at Eq.~(\ref{identity}). Plugging this relationship into the Eq.~(\ref{eq:TJNR}) leads to
\begin{align}
\begin{split}
\tjn &= \int_0^1 T(\phi)d\phi\\
&= \frac{T_0}{2}\left[-1 + \left(\frac{1}{\beta}+\beta\right)\text{arctan}(\beta)\right], \text{ } \beta = \frac{V}{2T_0\sqrt{L_0}}.
\end{split}
\end{align}

This final result is equivalent to Eq.\ (5.5) in Ref. \cite{sukhorukov_noise_1999} (differing only by a factor of 2 in the definition of $T_\textrm{JN}$) and reproduces Eq.\ (\ref{TJN1}) in the limit of small $V$.

\end{document}